\documentstyle[amssymb,aps,prl,floats,twocolumn,epsf]{revtex}
\begin{document}

\wideabs{
\title{Measurements of the Composite Fermion masses from the spin polarization of 2-D electrons in the region
$1<\nu<2$}
\author{R. Chughtai, V. Zhitomirsky, R.J. Nicholas}
\address{Department of Physics, Oxford University, \\Clarendon Laboratory, Parks Rd.,
Oxford, OX1 3PU, U.K.}
\author{M. Henini}
\address{School of Physics and Astronomy, Nottingham, NG7 2RD,U.K.}
\date{\today }
\maketitle

\begin{abstract}
Measurements of the reflectivity of a 2-D electron gas are used to
deduce the polarization of the Composite Fermion hole system
formed for Landau level occupancies in the regime $1<\nu<2$.  The
measurements are consistent with the formation of a mixed spin CF
system and allow the density of states or `polarization' effective
mass of the CF holes to be determined.  The mass values at $\nu=3/2$ are found to
be $\sim1.9m_{e}$ for electron densities of $4.4 \times 10^{11}
$cm$^{-2}$, which is significantly larger than those found from
measurements of the energy gaps at finite values of effective
magnetic field.
\end{abstract}

}

The fractional quantum Hall effect has been very successfully
described in terms of the picture of Composite
Fermions\cite{Jain}\cite{Halperin}, however considerable
uncertainties exist about both the Composite Fermion masses and
the role of spin and the formation of partially polarized states
\cite{Park,Du,Leadley,kukush1,Kronmuller}. In particular Park and
Jain \cite{Park} have pointed out that the CF mass measured from
activation measurements of the energy gap may be substantially
smaller than the thermodynamic or ``polarization'' mass which
determines the equilibrium spin populations of the carriers, an
idea which has received some recent experimental support
\cite{kukush1}. In this paper we describe a direct measurment of
the polarization of the 2D Composite Fermion (CF) gas and deduce
values for the polarization mass.

In the Composite Fermion model 2D electrons at filling factor $\nu=1/2$ form a new
collective state with quasi-particle excitations, known as Composite Fermions, and zero
effective magnetic field. If the Zeeman energy is small compared to the equivalent Fermi
energy in the Composite Fermion system the spin polarization of the 2D electrons around
$\nu=1/2$ deviates from 100$\%$ even at zero temperature. Cyclotron-like quantization of
the energy spectra of the Composite Fermions away from $\nu=1/2$ causes oscillations in
the spin polarization of the 2D electrons at integer filling factors for Composite
Fermions. Simultaneously, transport properties of the 2DEG become controllable by the
ratio of the cyclotron energy for the Composite Fermions to the Zeeman
energy\cite{Du}\cite{Eisenstein}. In spite of this evidence for partial polarization of
the CF system direct measurements of the CF Fermi wavevector suggest that the system is
close to fully polarized both at $\nu=1/2$ \cite{willett1} and at
$\nu=3/2$\cite{willett3}.

In this work we report direct measurements of the properties of
Composite Fermion holes in the region of level occupancy
$1<\nu<2$. For small values of the Zeeman energy a very convenient
way to treat the unoccupied states in the zeroth Landau level in
the region $1<\nu<2$ is as ``holes'' \cite{Du} with an effective
filling factor given by $\nu_{h} = 2 - \nu_{e}$ relative to
completely filled electron sublevels of both spins. The properties
of such holes in the region $1<\nu_{e}<2$ are equivalent to the
properties of electrons in the region $0<\nu_{e}<1$ if the energy
scale for the Coulomb interaction is much less than the cyclotron
energy. Thus a new Composite Fermion system is generated around
filling factor $\nu_{h}= 1/2$ which corresponds to the filling
factor $\nu_{e} = 2 - 1/2 = 3/2$. One should expect similar but
much more significant oscillations in the spin polarization of the
2D holes away from the filling factor $\nu_{e}$ = 3/2 \cite{Du} as
the relative energy of the Zeeman splitting is considerably
smaller with respect to the Coulomb interaction. Evidence of such
oscillations has been observed before\cite{kukush1} in the
magnetic field dependence of the spin polarization of electrons at
low density on the top of a strong monotonic fall of the electron
spin polarization towards $\nu = 2$. In principle there is a
simple relation between the spin polarization $P_{e}$ of 2D
electrons and the spin polarization $P_{h}$ of 2D holes
$P_{h}\nu_{h} = P_{e}\nu_{e}$, but in the region close to
$\nu_{e}=2$ both $\nu_{h}$ and $P_{e}$ become too small to allow
reliable extraction of the hole spin polarization from the
measured electron value. Here we measure directly the spin
polarization of the holes in the zeroth Landau level of 2D
electrons by measuring the hole population with interband
reflectivity. At temperatures below 1.5K we find that the spin
polarization of the holes in the region $1<\nu<2$ agrees very well
with a picture based on Composite Fermions with a relatively large
effective mass.

We have measured magnetoreflectivity from a 2DEG formed in a pair
of doped GaAs single quantum wells with widths of $120$\AA{} and
$150$\AA{} and with carrier densities of $4.4 \times 10^{11}$
cm$^{-2}$. The quantum wells were grown using a $300$\AA{} thick
Al$_{0.33}$Ga$_{0.67}$As spacer followed by a $300$\AA{} thick
layer of doped Al$_{0.33}$Ga$_{0.67}$As:Si with a doping level
$10^{18}$ cm$^{-3}$ and a final $250$\AA{} thick cap layer of
GaAs. The measurements used a standard projection lamp giving
unpolarized illumination of the sample in a 20T He$^{3}$ insert
with a base temperature of 0.4K. Reflected light from the sample
was analysed using an in-situ circular polariser and detected and
dispersed with a CCD camera and monochromator.

\begin{figure}[tbp]
\begin{center}
\epsfxsize=2.5in \epsffile{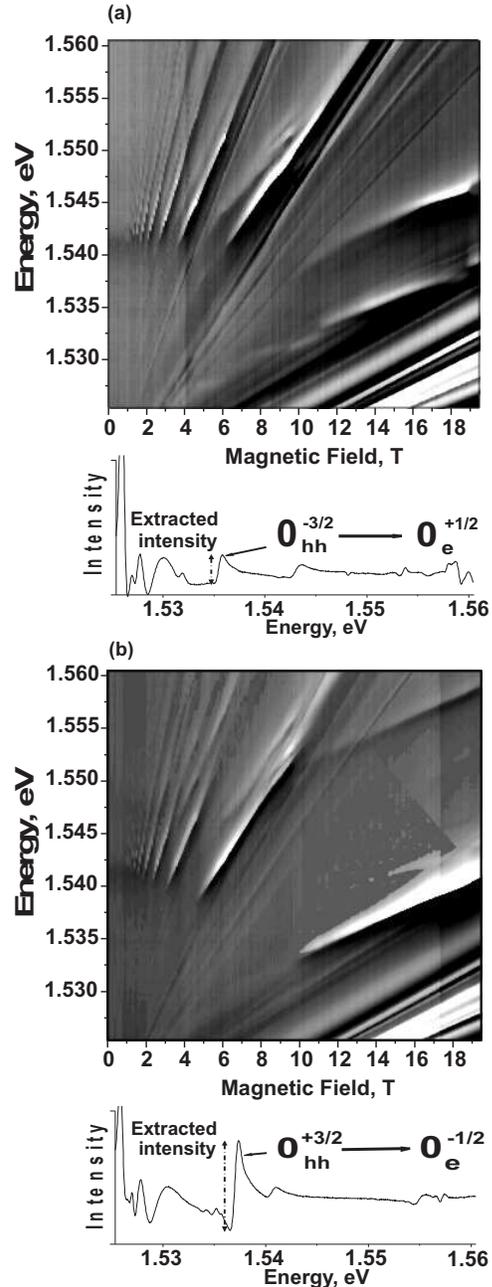} \caption[figure 3]{Upper section shows false
colour plots of reflectivity for the $\sigma^{-}$(a) and $\sigma^{+}$(b) polarizations.
Some LH transitions are also visible with opposite polarization dependence. The lower
sections show typical spectra measured at 14.5 T which illustrate the magnitude of the
reflectivity feature as used in the subsequent analysis.} \label{fig:spectra}
\end{center}
\end{figure}

Fig.1 shows a false colour plot of the optical reflectivity for both $\sigma^{+}$ and
$\sigma^{-}$ polarizations for the $150$\AA{} SQW. This is made from a series of
reflectivity spectra measured with a small magnetic field step at 0.4K. White corresponds
to high intensities of the reflected signal and black to a weaker signal. The strength of
the magnetoreflectivity signal is determined by the density of unoccupied states and
hence the density of holes in the Landau levels.  The absorption coefficient is directly
related to dispersion in the refractive index through the Kramers-Kronig relation and it
has been shown theoretically that the magnitude of the excursion of the reflectivity
around the transition energy is proportional to the absorption coefficient
\cite{Ivchencko}. The direct reflectivity spectra measured at 14.5T are shown as an
example of this analysis in the lower part of Fig.1.  At high magnetic fields the
reflectivity shows a set of different lines corresponding to the allowed transitions
between Landau levels of the valence and conduction bands with the same Landau level
index N. For a well resolved spin splitting in the 2D electron gas we start to detect N-N
transitions only after empty spaces appear in the Nth electron Landau level at filling
factor $\nu=2N+2$ for the upper spin sublevel ($\sigma^{+}$ polarization) and at
$\nu=2N+1$ for the lower spin sublevel ($\sigma^{-}$ polarization). This simple single
particle behaviour however changes dramatically for absorption into the two spin
sublevels of the zeroth Landau level. For $\nu<2$ at first we see only transitions to the
upper spin state, but beyond $\nu=5/3$ we start to detect transitions to both spin
sublevels in contrast to the predictions of a single particle picture.

\begin{figure}[tbp]
\epsfxsize=3.3in \epsffile{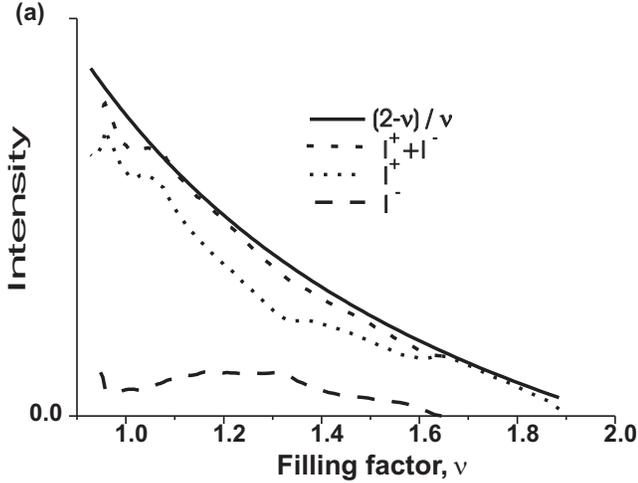} \vskip.02in\caption{The
intensities deduced from the reflectivity as a function of
occupancy at 0.4K for the $150$\AA{} quantum well.}
\label{fig:intensities}
\end{figure}
In Fig.2 we show the filling factor dependence of the relative
intensities $(I^+)$ and $(I^-)$ of the two lines from the
interband $0\downarrow\;$--$\;0\downarrow$ and
$0\uparrow\;$--$\;0\uparrow$ transitions extracted from the
magnetoreflectivity spectra for $\sigma^{+}$ and $\sigma^{-}$
polarized light. The sum of the intensities $(I^{+}+I^{-})$ for
both lines agrees well with the the theoretically expected
proportionality to $(2-\nu)/\nu$ for the density of empty states
thus confirming the validity of our analysis procedure. We
estimate from this that the relative values of the intensities are
correct to within $\pm5\%$. The spin polarization of the holes in
the zeroth Landau level should be equal to the polarization of the
intensities of the interband transitions
$P_{h}=(I^{+}-I^{-})/(I^{+}+I^{-})$. In Fig.3 we show $P_{h}$ for
the zeroth Landau level for temperatures in the range from $2.7$K
to $0.4$K. At temperatures below $1.5$K we find that the $P_{h}$
agrees very well with a picture based on Composite Fermions with a
g-factor close to that of the (bare) free electron in GaAs, as
shown schematically in the inset to Fig.3. We observe a partially
polarized flat region around filling factor $\nu_{h} =1/2$
$(\nu_{e} = 3/2)$ which corresponds to zero effective magnetic
field in the Composite Fermion picture. Moving into finite
effective magnetic field we observe a polarization approaching 1
at $\nu_{h} \leq1/3$ $(\nu_{e} \geq 5/3)$ which corresponds to the
complete or partial filling of only a single Composite Fermion
hole level and we find a well-developed minimum of the spin
polarization at $\nu_{h} =2/3$ $(\nu_{e} = 4/3)$ corresponding to
the filling of two oppositely oriented Composite Fermion hole
levels.

\begin{figure}[tbp]
\epsfxsize=3.3in \epsffile{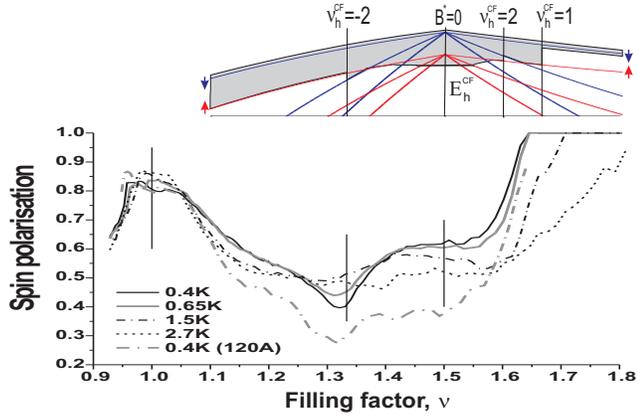} \vskip.02in\caption{Hole spin polarization as a
function of occupancy for several different temperatures for a 150\AA{} SQW. Also shown
is 0.4K data for the 120\AA{} SQW. Inset shows a schematic picture of the CF hole Landau
levels.} \label{fig:cflevels}
\end{figure}

From the absolute spin polarization at $\nu_{h} = 1/2$ we are able
to deduce the density of states effective mass (or polarization
mass \cite{Park}) of the Composite Fermions at zero effective
field from simple Pauli paramagnetism arguments. We assume that
the population difference is caused by the single particle (bare)
Zeeman energy, $g^{*}\mu_{B}B$, and that the level occupancy is
related to the spin polarization of the Composite Fermion holes by
$P_{h}=(\nu_{h}^{+}-\nu_{h}^{-})/(\nu_{h}^{+}+\nu_{h}^{-})$. If we
assume a constant CF mass (parabolic dispersion) we can relate the
effective Fermi energies for the spin up and spin down CF holes
$(E^{CF+}_f,\, E^{CF-}_f)$ to their occupancies and write

\begin{displaymath}
P_{h}=(E_{f}^{CF+}-E_{f}^{CF-})/(E_{f}^{CF+}+E_{f}^{CF-})
\end{displaymath}
\begin{displaymath}
 =g^{*}\mu_{B}B/(E_{f}^{CF+}+E_{f}^{CF-})
 \end{displaymath}
 \begin{equation}
 =(m^{*}g^{*}\mu_{B}B)/(2\pi\hbar^{2}n_{h})=m^{*}g^{*}/m_{e}
 \label{eqn:polariz}
\end{equation}

where $n_{h} = \frac{2eB}{h} - n_{e}$ is the number of CF holes.
The fact that we are using a parabolic dispersion means that the
same equation would hold for the electron polarization $P_{e}$ in
terms of the electron density $n_{e}$ and would lead to the same
values of mass for CF electrons and holes.

Using the measured $60\%$ hole spin polarization and remembering
that at $\nu_{e}=3/2$, $n_{h}=\frac{1}{3} n_{e}$ we find that
$m^*g^*/m_e = 0.60$ and the ratio $n_{h}^{+}/n_{h}^{-}=4$. The
electron g-factor depends on both the quantum well confinement and
the magnetic field and can be calculated using {\bf k.p} theory
\cite{sst} which may be approximated by the following expression:

\begin{equation}
g=2-20\left(\frac{1}{1.5+\Delta{E}_{g}}-
\frac{1}{1.85+\Delta{E}_{g}}\right)+0.062 \label{eqn:pres}
\end{equation}

where $\Delta{E}_g$ accounts for the increase in the band gap due to the electron and
hole confinement and cyclotron energies. This expression gives a g-factor of -0.32 for
the $150$\AA{} quantum well which gives a value of $m^{*}_{CF} \sim 1.9m_{e}$ for the CF
holes. Fig.3 also shows the spin polarization of the holes for a second $120$\AA{} SQW,
which has a significantly smaller g-factor of -0.23 (because of stronger confinement) but
the same carrier density. This sample shows a similar qualitative behaviour around
$\nu_{e} = 3/2$ ($\nu_{h} = 1/2$) but with a lower absolute polarization at $\nu=3/2$ of
only 0.40 due to the lower effective g-factor.  The polarization value gives a similar
value for the effective mass of $1.74m_{e}$ and confirms the general picture.

We can further confirm the large value of the effective mass by
examining the temperature dependence of the spin polarization of
the CF holes moving in finite effective field. The total $(B)$ and
effective $(B^*)$ fields for the CF hole levels are given by
\begin{equation}
B=\frac{2p+1}{3p+2} n_{e}\Phi_{0}\quad  \textrm{and,}\quad
B^{*}=\frac{1}{3p+2} n_{e}\Phi_{0}
 \label{eqn:fields}
\end{equation}
where $p =\pm1,2,\ldots,\, \Phi_{0}$ is the flux quantum and for
the present samples $n_{e}\Phi_{0}=18$T. For the case of $p=\pm2$
when the Zeeman and CF cyclotron energies are comparable we expect
a high temperature limit for the polarization of $50\%$, as there
will be $1\frac{1}{2}$ filled levels of spin up and a
$\frac{1}{2}$ filled level of spin down. The low temperature
behaviour will now depend on which splitting is largest with
$g^*\mu_B B>\hbar eB^*/m^*$ leading to an increase in polarization
as $T\rightarrow0$ and $g^*\mu_B B<\hbar eB^*/m^*$ leading to a
falling polarization. The low temperature limits for the two cases
should be 100\% and 0\% for totally unbroadened levels. The actual
values will be less than these limits due to the disorder present
and the fact that the lowest temperature studied is only 0.4K.
However, the direction of the temperature dependence will be
independent of the disorder unless this is highly asymmetric. For
$p=-2$ $(\nu_h=2/3,\, \nu_e=4/3)$ and $p=+2$ $(\nu_h=2/5,\,
\nu_e=8/5)$ the high $T$ limit (2.7K) is $P=50\%$ as expected but
for $p=-2$ the polarization falls to $\sim 40\%$ and for $p=+2$ it
rises to $\sim 75\%$. These two cases give us the limits
$1.2<m^*<2.0$, with the fact that the polarization at $p=+2$ is
closer to its fully polarized value suggesting that the mass is
closer to the upper limit. Around $\nu_e=3/2$ $(\nu_h=1/2)$ there
is a rapid depolarization of the spin system with temperature due
to the increasing thermal population of the upper spin sublevel,
which is consistent with the behaviour being dominated by the
small energy scale of the bare spin splitting. By contrast at
$\nu_{e} = 1$ $(\nu_{h} =1)$ there is a maximum in the spin
polarization of the holes which is much more robust than these
other features due to the formation of the Quantum Hall
ferromagnet and exchange enhancement of the spin gap, although the
polarization is not $100\%$ due to spin mixing phenomena which
will be the subject of a further publication \cite{zhit}.

The values which we have deduced for the mass are high compared with values based on
activation measurements of the energy gap \cite{CFmass,Du2,lkmass}, exactly as predicted
by Park and Jain \cite{Park}, and it should be born in mind that the value of the carrier
density is quite high in the samples studied here and we would expect that the mass would
increase due to the $\sqrt{B}$ dependence of the Coulomb energy on magnetic field as
observed in activation measurements for $\nu_e<1$ \cite{CFmass}. The value for
$m^*g^*/m_e$ is also substantially higher than found from tilted field \cite{Du,Gee,Du3}
and thermal acivation\cite{Du3} measurements around $\nu = 3/2$ in lower density samples.
Du et al \cite{Du} in particular concluded that there was a dependence on the effective
field of $m^*g^*/m_e = 0.264 + 0.05B^*$ which would give a value of $m^*g^*/m_e = 0.49$
for $\nu=4/3$ in our samples. Similarly the conclusion of Leadley et al \cite{CFmass} was
that the mass alone also depended on the effective field as $m^*/m_e = 0.51 + 0.083B^*$
which would give a mass of $1.91m_e$ at $\nu_e = 1/3$ in our samples. The present
measurements suggest instead that the polarization mass is determined only by the total
carrier density and the fact that the polarization approaches $\nu=3/2$ smoothly shows
that there is no evidence for any divergence of the mass as $B^* \rightarrow 0$ as has
been previously suggested in both theory \cite{Halperin} and experiment\cite{Du4}. One
question is whether the $\nu_{h}=1/2$ CF hole masses should be compared with the mass at
the equivalent (low) electron densities or with the value appropriate to the total
electron density. In the present case the parabolic dispersion assumption implies that
the CF hole and electron masses should be the same and therefore makes it likely that the
high values for the CF hole mass are due to the high total carrier density. Even higher
values for the masses have been deduced in a recent optical measurement of electron
polarization deduced indirectly from photoluminescence polarization by Kukushkin et al
\cite{kukush1}. Values of $2.27 m_{e}$ were found both at $\nu_{e}=1/2$ with $n_{e}=1.26
\times 10^{11}$cm$^{-2}$ and at $\nu_{e}=3/2$ with $n_{e}=6.3 \times 10^{11}$cm$^{-2}$.
The higher density result seems consistent with our data, but not that at the lower
density.

In conclusion, we have demonstrated that a direct measurement of
the spin polarization of the 2DEG can be made through measurements
of the absorption deduced from the reflectivity. This allows us to
observe a partially polarized CF gas and to deduce the
polarization mass of the CF holes which is found to be
significantly higher than the values deduced from measurements of
the activation energy gaps at fractional filling factors.

{\bf Acknowledgements:} \ We are grateful to the UK-EPSRC for
continued support.

\end{document}